\def\paperversiondraft{draft}

\ifx\paperversion\paperversiondraft
\else
  \def\paperversion{normal}
\fi

\def\grammarlyon{on}

\ifx\grammarly\grammarlyon
\def\ClassReview{}
\else
\def\ClassReview{review,}
\fi

\documentclass[\ClassReview nonacm, sigplan, 10pt, pageno, review=false]{acmart}

\usepackage[frozencache,cachedir=.]{minted}

\ifx\grammarly\grammarlyon
\usepackage[none]{hyphenat}
\else
\fi

\usepackage{multicol}
\usepackage{tabu}

\usepackage{colortbl}
\usepackage{wrapfig}
\usepackage{diagbox}

\usepackage{environ}
\ifx\paperversion\paperversiondraft

\else
\NewEnviron{draftonly}{}
\fi

\usepackage[utf8]{inputenc}
\usepackage[T1]{fontenc}
\usepackage{microtype}
\usepackage{amsmath}
\usepackage{graphicx}
\usepackage{caption}
\usepackage{subcaption}
\usepackage{listings}
\usepackage{wrapfig}
\usepackage{natbib}

\makeatletter
\let\FV@ListProcessLineOrig\FV@ListProcessLine
\def\FV@ListProcessLine#1{%
  \ifx\FV@Line\empty
    \hbox{}\vspace{\dimexpr-\baselineskip+1.5\smallskipamount}%
  \else
    \FV@ListProcessLineOrig{#1}%
  \fi}
\makeatother

\usepackage{xargs}
\usepackage{lipsum}
\usepackage[textsize=tiny]{todonotes}
\usepackage{xparse}
\usepackage{xifthen, xstring}
\usepackage[normalem]{ulem}
\usepackage{xspace}
\usepackage{marginnote}

\usepackage{pifont}

\makeatletter
\renewcommand{\paragraph}{%
  \@startsection{paragraph}{4}%
  {1ex}{0ex \@plus 1ex \@minus .2ex}{-.5em}%
  {\normalfont\normalsize\bfseries}%
}
\makeatother

\NewDocumentCommand{\code}{v}{%
\texttt{#1}%
}

\NewDocumentCommand{\xdslcode}{v}{%
\texttt{#1}%
}

\NewDocumentCommand{\pycode}{v}{%
\mintinline{python}{#1}%
}

\makeatletter
\patchcmd{\@addmarginpar}{\ifodd\c@page}{\ifodd\c@page\@tempcnta\m@ne}{}{}
\makeatother
\ifx\paperversion\paperversiondraft
  \makeatletter
    \paperwidth=\dimexpr \paperwidth + 6cm\relax
    \oddsidemargin=\dimexpr\oddsidemargin + 3cm\relax
    \evensidemargin=\dimexpr\evensidemargin + 3cm\relax
    \marginparwidth=\dimexpr \marginparwidth + 3cm\relax
  \makeatother
\fi

\usepackage[textsize=tiny]{todonotes}
\usepackage[normalem]{ulem}

\makeatletter
\font\uwavefont=lasyb10 scaled 652
\newcommand\colorwave[1][blue]{\bgroup\markoverwith{\lower3\p@\hbox{\uwavefont\textcolor{#1}{\char58}}}\ULon}

\newcommand\highlight[2]{{{\colorwave[#1]{#2}}}}
\makeatother

\makeatletter
\newcommand\InFloat[2]{\ifnum\@floatpenalty<0\relax#1\else#2\fi}
\makeatother

\ifx\paperversion\paperversiondraft
\newcommand\createtodoauthor[2]{
  \def\tmpdefault{emptystring}
  \expandafter\newcommand\csname #1\endcsname[2][\tmpdefault]{
    \def\tmp{##1}
    \InFloat{
        \smash{
	  \marginnote{
	    \todo[inline,linecolor=#2,backgroundcolor=#2,bordercolor=#2]
	      {\textbf{#1 (Figure):} ##2}
          }
        }
    }{\ifthenelse{\equal{\tmp}{\tmpdefault}} 
      {\todo[linecolor=#2,backgroundcolor=#2,bordercolor=#2]{\textbf{#1:} ##2}\ignorespaces}
      {\ifthenelse{\equal{##2}{}} 
        {\highlight{#2}{##1}}
        {\highlight{#2}{##1}\todo[linecolor=#2,backgroundcolor=#2,bordercolor=#2]
	  {\textbf{#1:} ##2}
	}
      }
    }
  }
}
\else
\newcommand\createtodoauthor[2]{%
\expandafter\newcommand\csname #1\endcsname[2][]{##1}%
}%
\fi

\usepackage{minted}
\usepackage{etoolbox}

\makeatletter
\ifcsdef{minted@optlistcl@quote}
{
\ifwindows
  \renewcommand{\minted@optlistcl@quote}[2]{%
    \ifstrempty{#2}{\detokenize{#1}}{\detokenize{#1="#2"}}}
\else
  \renewcommand{\minted@optlistcl@quote}[2]{%
    \ifstrempty{#2}{\detokenize{#1}}{\detokenize{#1='#2'}}}
\fi

\newcommand{\minted@def@optcl@novalue}[2]{%
  \define@booleankey{minted@opt@g}{#1}%
    {\minted@addto@optlistcl{\minted@optlistcl@g}{#2}{}%
     \@namedef{minted@opt@g:#1}{true}}
    {\@namedef{minted@opt@g:#1}{false}}
  \define@booleankey{minted@opt@g@i}{#1}%
    {\minted@addto@optlistcl{\minted@optlistcl@g@i}{#2}{}%
     \@namedef{minted@opt@g@i:#1}{true}}
    {\@namedef{minted@opt@g@i:#1}{false}}
  \define@booleankey{minted@opt@lang}{#1}%
    {\minted@addto@optlistcl@lang{minted@optlistcl@lang\minted@lang}{#2}{}%
     \@namedef{minted@opt@lang\minted@lang:#1}{true}}
    {\@namedef{minted@opt@lang\minted@lang:#1}{false}}
  \define@booleankey{minted@opt@lang@i}{#1}%
    {\minted@addto@optlistcl@lang{minted@optlistcl@lang\minted@lang @i}{#2}{}%
     \@namedef{minted@opt@lang\minted@lang @i:#1}{true}}
    {\@namedef{minted@opt@lang\minted@lang @i:#1}{false}}
  \define@booleankey{minted@opt@cmd}{#1}%
      {\minted@addto@optlistcl{\minted@optlistcl@cmd}{#2}{}%
        \@namedef{minted@opt@cmd:#1}{true}}
      {\@namedef{minted@opt@cmd:#1}{false}}
}
\minted@def@optcl@novalue{customlexer}{-x}
}
{
}
\makeatother

\makeatletter
\ifcsdef{minted@optlistcl@quote}
{
\newminted[mlir]{tools/MLIRLexer.py:MLIRLexerOnlyOps}{customlexer, mathescape}
\newminted[xdsl]{tools/MLIRLexer.py:MLIRLexer}{customlexer, mathescape, style=murphy}

}
{
\newminted[mlir]{tools/MLIRLexer.py:MLIRLexerOnlyOps -x}{mathescape}
\newminted[xdsl]{tools/MLIRLexer.py:MLIRLexer -x}{mathescape, style=murphy}
}
\makeatother
\newminted{python}{fontsize=\footnotesize,escapeinside=||}

\graphicspath{{./images/}}

\makeatletter
\newcommand\requiredelimiter[2][########]{%
  \ifdefined#2%
    \def\@temp{\def#2#1}%
    \expandafter\@temp\expandafter{#2}%
  \else
    \@latex@error{\noexpand#2undefined}\@ehc
  \fi
}
\@onlypreamble\requiredelimiter
\makeatother

\newcommand\newdelimitedcommand[2]{
\expandafter\newcommand\csname #1\endcsname{#2}
\expandafter\requiredelimiter
\csname #1 \endcsname
}

\usepackage{booktabs}

\usepackage{array}
\newcolumntype{P}[1]{>{\centering\arraybackslash}p{#1}}

\newdelimitedcommand{toolname}{\textit{xDSL}}
\newdelimitedcommand{nTotalOps}{1862}
\newdelimitedcommand{nTotalMlirOps}{1554}
\newdelimitedcommand{nTotalToolNameOps}{296}
\newdelimitedcommand{nMeanOps}{27}

\newdelimitedcommand{nTestsInFolders}{4868}
\newdelimitedcommand{nTestsInFoldersPassing}{4681}
\newdelimitedcommand{nTestsInFoldersFailing}{187}
\newdelimitedcommand{percentageTestsInFoldersPassed}{96.2\%}

\newdelimitedcommand{suiteParsingTimeMLIRDebug}{29}
\newdelimitedcommand{suiteParsingTimeMLIRRelease}{6}
\newdelimitedcommand{suiteParsingTimexDSL}{1029}
\newdelimitedcommand{suiteParsingTimexDSLNative}{34}

\usepackage{tikz}
\usetikzlibrary{arrows}
\usetikzlibrary{shapes}
\DeclareRobustCommand\circled[2][]{\ifthenelse{\isempty{#1}}{\tikz[baseline=(char.base)]{\node[shape=circle,fill=pairedOneLightBlue,inner sep=1pt] (char) {#2};}}{\autoref{#1}: \hyperref[#1]{\tikz[baseline=(char.base)]{\node[shape=circle,fill=pairedOneLightBlue,inner sep=1pt] (char) {#2};}}}}
\DeclareRobustCommand\circle[2][]{\tikz[baseline=(char.base)]{\node[shape=circle,fill=#1,inner sep=1pt] (char) {\color{#1}#2}}}

\setcopyright{none}             
\usepackage[most]{tcolorbox}
\tcbuselibrary{minted}
\definecolor{outcolor}{HTML}{D84315}
\definecolor{Incolor}{HTML}{D84315}

\newlength{\promptwidth}
\newlength{\promptsep}
\newcommand{\prompt}[4]{%
    \makebox[0pt][r]{\texttt{\color{#2}#1[#3]:#4}}\vspace{-1.25\baselineskip}%
}

\newcounter{NBout}

\newtcblisting{NotebookIn}{
    breakable,
    boxrule=0pt,
    size=fbox,
    left skip = \promptwidth + \promptsep,
    pad at break*=5mm,
    opacityfill=1,
    colback=black!5,
    phantom=\refstepcounter{NBout},
    title=\prompt{In}{black}{\theNBout}{\hspace{\promptsep}},
    fonttitle=\linespread{1}\tiny,
    attach title to upper,
    listing only,
    minted language = python,
    minted options={fontsize=\tiny},
    boxrule=0pt,
    arc=0pt,
    frame hidden,
    colframe=white!75!white,
}

\newtcblisting{NotebookOut}{
    breakable,
    boxrule=0pt,
    size=fbox,
    left skip = \promptwidth + \promptsep,
    pad at break*=5mm,
    opacityfill=0,
    title=\prompt{Out}{black}{\theNBout}{\hspace{\promptsep}},
    fonttitle=\linespread{1}\tiny,
    attach title to upper,
    listing only,
    minted language = text,
    minted options={fontsize=\tiny},
}

\begin{document}

\title{Sidekick compilation with \toolname{}}
\date{}

\author{Mathieu Fehr}
\email{mathieu.fehr@ed.ac.uk}
\affiliation{
  \institution{The University of Edinburgh}
  \country{}
}

\author{Michel Weber}
\email{michel.weber@inf.ethz.ch}
\affiliation{
  \institution{ETH Zurich}
  \country{}
}

\author{Christian Ulmann}
\email{christian.ulmann@inf.ethz.ch}
\affiliation{
  \institution{ETH Zurich}
  \country{}
}

\author{Alexandre Lopoukhine}
\email{lopoukhine@ed.ac.uk}
\affiliation{
  \institution{The University of Edinburgh}
  \country{}
}

\author{Martin Lücke}
\email{martin.luecke@ed.ac.uk}
\affiliation{
  \institution{The University of Edinburgh}
  \country{}
}

\author{Théo Degioanni}
\email{theo.degioanni@ens-rennes.fr}
\affiliation{
  \institution{ENS Rennes}
  \country{}
}

\author{Michel Steuwer}
\email{michel.steuwer@tu-berlin.de}
\affiliation{
  \institution{Technische Universität Berlin}
  \country{}
}

\author{Tobias Grosser}
\email{tobias.grosser@cst.cam.ac.uk}
\affiliation{
  \institution{University of Cambridge}
  \country{}
}

\begin{abstract}
Traditionally, compiler researchers either conduct experiments within an existing
production compiler or develop their own prototype compiler; both options come
with trade-offs. On one hand, prototyping in a production compiler can be
cumbersome, as they are often optimized for program compilation speed at the
expense of software simplicity and development speed. On the other hand, the
transition from a prototype compiler to production requires significant
engineering work.
To bridge this gap, we introduce the concept of \textbf{sidekick compiler
frameworks}, an approach that uses multiple frameworks that interoperate with
each other by leveraging textual interchange formats and declarative
descriptions of abstractions. Each such compiler framework is specialized for
specific use cases, such as performance or prototyping.
Abstractions are by design shared across frameworks, simplifying
the transition from prototyping to production.
We demonstrate this idea with \toolname{}, a sidekick for MLIR focused on
prototyping and teaching. \toolname{} interoperates with MLIR through a shared
textual IR and the exchange of IRs through an IR Definition Language.
The benefits of sidekick compiler frameworks are evaluated by showing on three use
cases how \toolname{} impacts their development: teaching, DSL compilation,
and rewrite system prototyping. We also investigate the trade-offs that
\toolname{} offers, and demonstrate how we simplify the transition between
frameworks using the IRDL dialect.
With sidekick compilation, we envision a future in which engineers minimize the
cost of development by choosing a framework built for their immediate needs, and
later transitioning to production with minimal overhead.

\end{abstract}

\maketitle

\section{Introduction}

When conducting compiler research, one can either embed a new idea in a
preexisting compiler or develop an entirely new prototype compiler intended to
show a speedup or other benefits. While writing a prototype may be more flexible
in the short term, it comes with several drawbacks. First, a new prototype
requires the reimplementation of many features, such as IR data structures, a
parser and printer for the IR textual format, or generic passes such as dead
code elimination. Later on, if the prototype is promising, porting it to a
production compiler requires significant work, often a complete rewrite, and
cannot always be done iteratively.

\begin{figure}
  \includegraphics[width=\columnwidth]{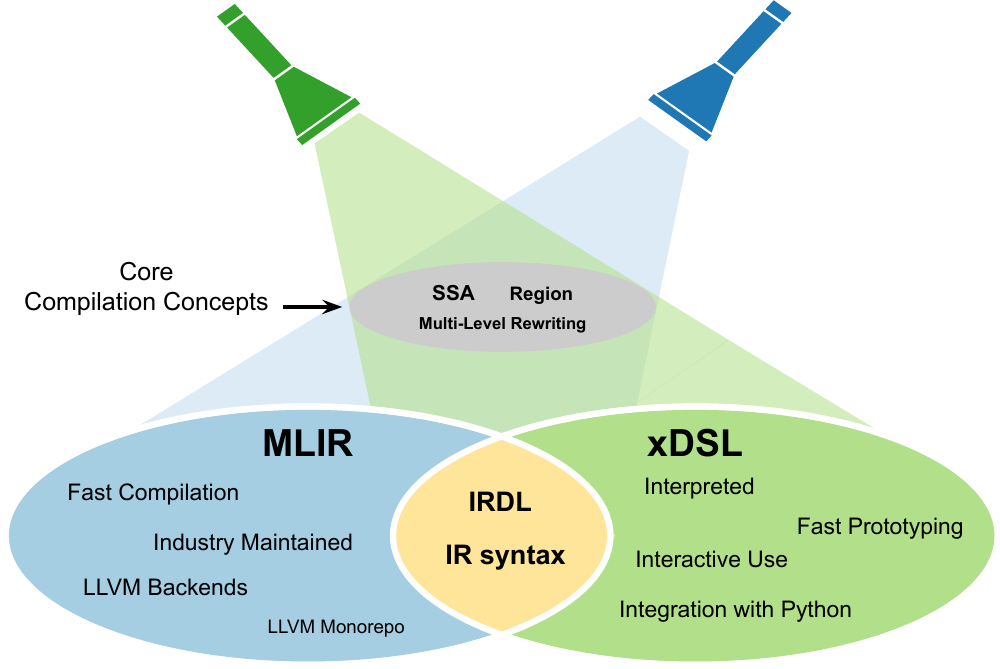}
  \caption{Compiler frameworks that share the same core compilation concepts can
  offer tailored projections of these concepts for specific use cases while
  being still compatible through IRs and declarative IR definitions.}
  \label{fig:abstract-overview}
\end{figure}

On the other hand, compiler frameworks reduce the cost of
working in a production compiler by providing modular infrastructure that can
easily be reused. For instance, MLIR~\cite{lattner2021mlir} allows users to
define their own abstractions, or to reuse abstractions from the ecosystem,
which covers machine learning kernels~\cite{vasilache22}, stencil
computations~\cite{gysi2021domain}, quantum computing~\cite{peduri22}, hardware
design~\cite{CIRCT}, and more. Due to the uniform representation that
MLIR offers, abstractions can be freely combined, enabling compilation flows
that combine domains, e.g., running ML inference in a database-style
query~\cite{LingoDB}. Consequently, compiler frameworks facilitate compiler
design and reduce prototyping costs.

However, while MLIR allows compiler experts to extend the compiler,
it requires users to work in a setting that is designed for compile-time and
runtime performance. As an industry-funded project that targets
performance-focused production use cases, MLIR follows LLVM in its choice of
C++ as its implementation language, benefiting from a fast implementation
deployable across many target systems.  Furthermore, MLIR's integration with
LLVM goes beyond the implementation language since IRs are defined using LLVM's
in-house TableGen language, LLVM's abstract data types are preferred over
standard C++, and MLIR uses several low-level performance optimizations,
e.g., to enable fast type equivalence checks via pointer comparison.  While all
these choices are justified, they mean that working with MLIR requires
expertise in C++ and LLVM and costly development recompilation cycles --
constraints that are hard to justify in some circumstances, in particular in
teaching and research.

In order to facilitate other use cases such as prototyping, not only abstractions need
to be connected in a modular way, but frameworks themselves. To that end, we
propose the idea of {\bf\emph{sidekick compiler frameworks}}, which are
frameworks loosely coupled through the use of shared core compilation concepts
(\autoref{fig:abstract-overview}) and aligned exchange formats for IRs and IR
definitions. In particular, a sidekick framework can be interleaved at any point
in the compilation pipeline with its base framework. By deliberately reusing the
same core compilation concepts, like SSA and multi-level rewriting, and
exchanging IR definitions between frameworks through a common format, a sidekick
framework eases the transition between frameworks.

In this paper, we present \toolname{} \footnote[1]{~~\href{https://github.com/xdslproject/xdsl}{https://github.com/xdslproject/xdsl}},
a sidekick compiler framework for MLIR written in
Python. \toolname{} is standalone, but interacts with MLIR through a shared
textual IR format and the IRDL~\cite{irdl} dialect, an MLIR-based meta-IR that expresses IR
definitions as programs. To evaluate the benefits of \toolname{} as a sidekick
compiler framework, we explore three use cases that benefit from
\toolname{}'s Python implementation and show several statistics to
compare \toolname{} and MLIR. Our analysis shows that bringing state-of-the-art
concepts of MLIR to new use cases result in a broader and better-connected
compiler ecosystem that can cater to various novel workflows.

\vspace{.5em}
\noindent
Our contributions are:
\begin{itemize}
  \item The concept of a \emph{sidekick compiler framework} coupled to a base framework via
	deliberately sharing compilation concepts
	(e.g., SSA-based IRs, nested regions, attributes) and compatible textual IRs
        (Sections \ref{sec:compiler} and \ref{sec:pykick}).
  \item Three case studies that characterize workflows that benefit
	from sidekick compilation (\autoref{sec:usecases})
\item An encoding of IR definitions as SSA-based compiler IR (the IRDL dialect) for the
      exchange of IR definitions between our sidekick and base framework.
  (\autoref{sec:dialect-definition})
  \item A comparison of several user-relevant metrics between \toolname{} and MLIR.
	(\autoref{sec:design_space})
\end{itemize}

\definecolor{figOneC1}{HTML}{9437ff}
\definecolor{figOneC2}{HTML}{009051}
\definecolor{figOneC3}{HTML}{941100}
\definecolor{figOneC4}{HTML}{ff9300}
\definecolor{figOneC5}{HTML}{0433ff}

\section{Sidekick Compilation}
\label{sec:compiler}

We define sidekick compiler frameworks as compiler frameworks that can be used in a completely
standalone way, yet can be connected to a base framework through a common
IR exchange format. To achieve this, sidekick frameworks use the same core compilation
concepts as the base framework, such as SSA-based IRs, or nested regions. A
fundamental advantage of sidekick frameworks is that their implementation
can be tailored to the needs of their target audience, e.g., by providing a
simple implementation of the core IR to ease prototyping, or by providing a
verified implementation to enable formal verification of the compiler. Finally,
sidekick frameworks can share IR definitions through an additional exchange format,
simplifying the porting of code between the two frameworks.

\begin{figure}
  \centering
  \includegraphics[width=17em]{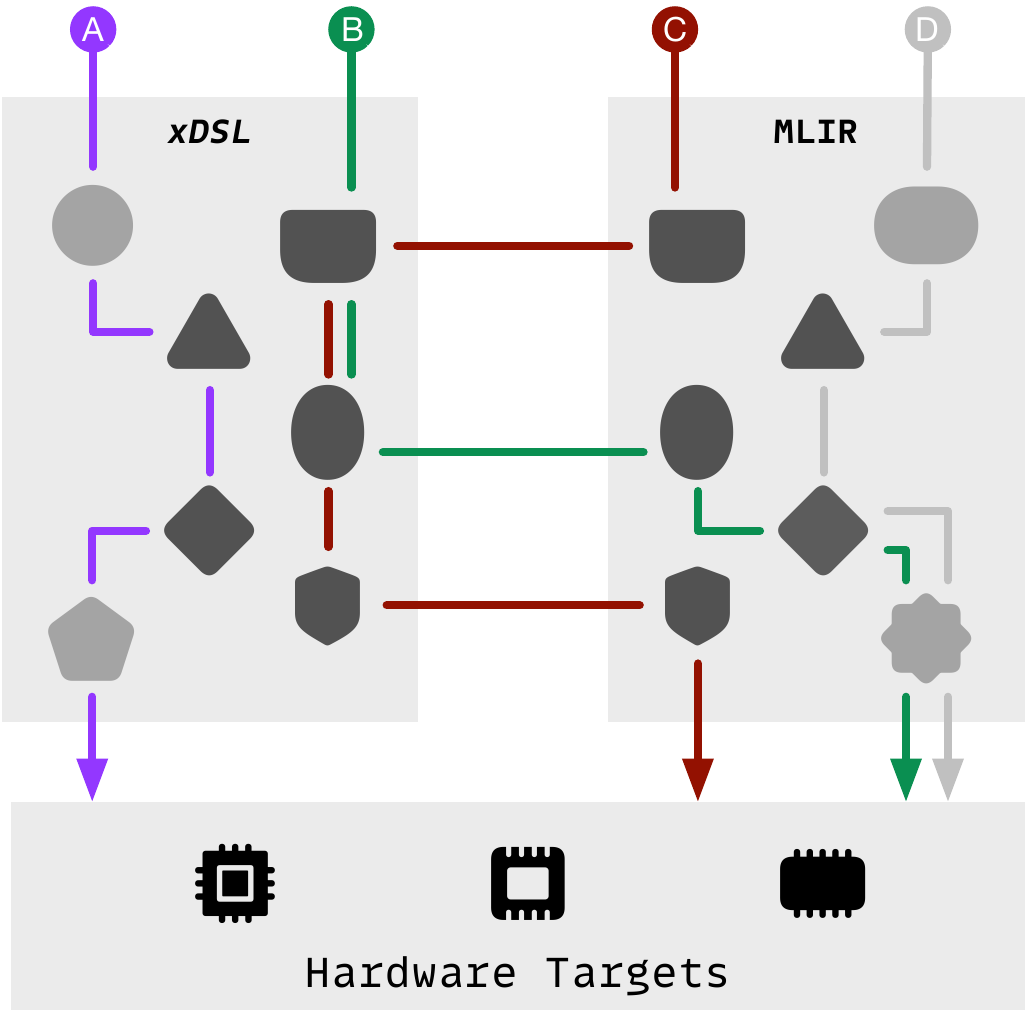}
  \vspace{1em}
  \caption{Dialect definitions (grey shapes in \toolname and MLIR) can be shared
  between \toolname{} and MLIR (dark-grey shapes exist in both). This enables the
  exploration of new workflows in the compiler space:
  \circle[figOneC1]{\color{white}\tiny A} python-native development of end-to-end
  compilers,
  \circle[figOneC2]{\color{white}\tiny B} pairing high-level
  DSLs with low-level compilers,
  and \circle[figOneC3]{\color{white}\tiny C} extending low-level compilers to
  explore new compiler design ideas. }
  \label{fig:xdsl-flow}
  \end{figure}

We demonstrate the concept of sidekick compilation with \toolname{}, a
Python-native compiler framework that interoperates with MLIR. \toolname{}
offers a standalone compiler framework that targets developers who use
Python for their main workflows, or want to quickly prototype new compiler
ideas or abstractions. The dynamic nature of Python also allows the use
of \toolname{} in new environments such as Jupyter Notebooks~\cite{kluyver2016jupyter}.
The core novelty of \toolname{} is its coupling with the base compiler MLIR
({\autoref{fig:xdsl-flow}}). \toolname{} couples to MLIR by mirroring
its core IR structure and its textual IR.

A sidekick compiler's ability to exchange IR definitions and programs with its
base compiler enables several new workflows, some of which we highlight
briefly. The most straightforward workflow
\circle[figOneC1]{\color{white}\footnotesize A} uses \toolname{} as a
standalone compiler framework to implement a self-contained compiler. The core
property of such a workflow is that it stays entirely within Python, enabling
developers to run the compiler on any Python-supported platform, iterate
quickly by extending the compiler at runtime, or integrate Python libraries
with ease into the compilation flow.  Interestingly, even an independently used
sidekick still can leverage IR definitions that were initially developed in
the base compiler. \toolname{} also enables workflows
\circle[figOneC2]{\color{white}\footnotesize B} that yield a full DSL
compiler by combining a domain-specific front-end implemented in Python with
the hardware targets available in MLIR. The use of MLIR and LLVM offers
additional low-level optimizations, powerful register allocation and
instruction selection, as well as infrastructure for targeting the latest hardware
accelerators. Furthermore, \toolname{} allows workflows
\circle[figOneC3]{\color{white}\footnotesize C} where \toolname{} is placed
into a pre-existing compilation flow of the baseline compiler. Such workflows
make it possible to prototype new compilation approaches and ideas from Python,
for example, new rewriting systems. By porting IR from MLIR to \toolname{}, using
the \toolname{}-based prototype, and then going back to MLIR, one can show the potential of
new approaches using an early Python-based prototype. While these examples
correspond to the use cases shown later (\autoref{sec:usecases}), they are not
exhaustive, and we expect other future uses.

One of the core novelties that enables sidekick compilation in \toolname{} is
its ability to share IR definitions with MLIR. Expanding on MLIR's declarative IR
definition language IRDL~\cite{irdl}, we made it possible to share IR definitions
between \toolname{} and MLIR by encoding IR definitions
with a novel SSA-based meta-IR, the IRDL dialect, that can be exchanged between compilers like any other program.
\toolname{} and MLIR can both translate their IR definitions into the IRDL dialect and
import IRDL dialect definitions to instantiate externally provided IRs.
Implementing this is relatively easy. Both \toolname{} and MLIR share the same
definition of the IRDL dialect meta IR. As we can translate programs between the
two frameworks, we can also translate IR definitions and, therefore, use
IRs defined in MLIR from \toolname{} and vice-versa. This gives access to the
existing ecosystem of MLIR dialects and also makes importing IRs from
\toolname{} into MLIR possible.

While a sidekick compiler such as \toolname{} does not share transformations
with MLIR, we are working towards connecting \toolname{} with MLIR's PDL
dialect, a dialect to define and reason about IR rewrites. Using PDL, we expect
to eventually be able to port rewrites from one compiler to the other,
making the rewrites themselves less implementation-dependent and potentially
allowing the transfer of increasingly complete compilation flows across compiler boundaries.

\section{Three Use Cases for SideKick Compilation}
\label{sec:usecases}

We demonstrate three use cases by discussing the users and
their respective needs, how existing workflows address, or fail to address,
those needs, and how \toolname{} facilitates the uses case or even enables them
in the first place. We used \toolname{} to teach compilation at the University of
Edinburgh, implement a DSL compiler that leverages MLIR's low-level
optimizations to reach state-of-the-art performance, and prototype a new
rewriting engine for the multi-level rewriting approach.

\subsection{Use case 1: Teaching compilation with ChocoPy}
\label{sec:uc1}

\begin{wrapfigure}[9]{r}{13em}
  \vspace{-2em}
  \center
  ~\includegraphics[width=11em]{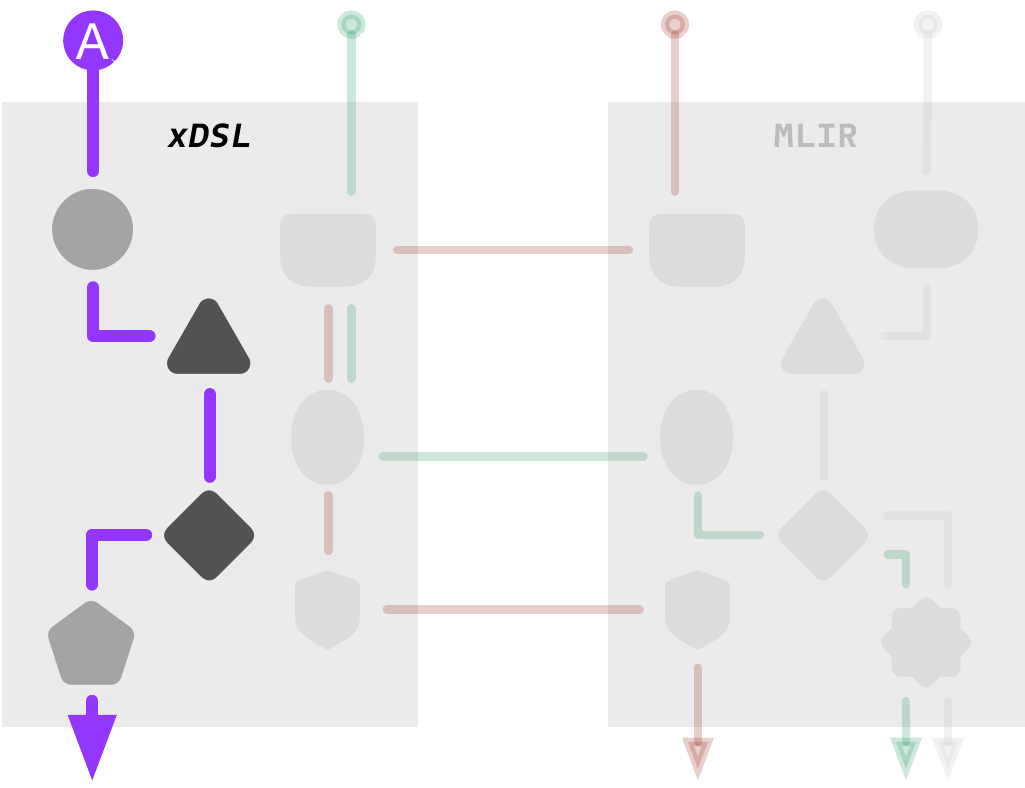}
  \caption{A \toolname{}-only flow.}
\end{wrapfigure}

While most compilers are written by professionals and for production use,
implementing a compiler is also a great way to teach compilation concepts.
However, writing an SSA-based compiler from scratch is a complex task, and using
frameworks helps focus on the compilation concepts rather than the
implementation details of the data structures in typical SSA compilers. We used
\toolname{} for two years at the University of Edinburgh compilation class to
teach around 200 students. We tasked students with implementing a compiler for
ChocoPy~\cite{padhye2019chocopy}, a subset of Python designed to teach
compilation to students. Students must implement a parser, a type checker,
optimizations at multiple abstraction levels, and a pass lowering the IR from
a middle-end IR to a RISCV IR.

\paragraph{\textbf{User}}

While typical compiler engineers often have high-end computers capable of
quickly building and running production-quality compilers, students may have
slower computers with diverse architectures and operating systems.
Students often have less programming experience and less experience with package
managers and build systems. Finally, most students, especially in introductory
classes, may not pursue compiler work in the future, and thus investing much
time in a framework may not be valuable for them.

\paragraph{\textbf{Needs}}

Students have different needs than typical compiler engineers. Because of consumer
hardware, compiling large frameworks is often impractical, and long
incremental building times are a big source of frustration. The installation
needs to be simple and quick since many students are unfamiliar with
build systems. Optimally, students should be able to install and start trying
out the framework during a single lab session (2 hours). Finally, frameworks
that are not portable across architectures and operating systems are known
sources of problems. A compiler implemented as a lecture project is not required
to be as fast as industry compilers. Thus, frameworks with longer compilation times
than state-of-the-art compilers are not an issue.

\paragraph{\textbf{Existing Workflows}}

One existing workflow would be to not provide any framework to the compiler
class and, instead, ask students to implement their own data structures. This
is, for instance, the approach taken by the original ChocoPy compiler class. We
argue that this solution is not optimal for the students, since they have to
implement a lot of boilerplate code that is unrelated to the compilation
techniques they are learning. Thus, the time required to set up an end-to-end
compilation flow already consumes most of the available time in a lecture
project setting. Also, most of these compiler classes usually do not teach the
important SSA representation because it is hard to write SSA compiler infrastructure.

Another possibility is using an existing compiler infrastructure. Some compiler
classes use LLVM, but LLVM can only be used as a mid-level IR and cannot be used
for high-level or low-level IRs. While MLIR has the core compilation concepts
that interest us, it is both complex and slow to install. Also, iterating on a
compiler is expensive with MLIR, especially with low-end machines. Finally, MLIR
has a steep learning curve due to its low-level nature, and thus students may
lose significant time understanding the framework.

\paragraph{\textbf{The \toolname{} Approach}}

We argue that using \toolname{} is the best solution for the students.
Students are able to install the framework and start using it in a single lab,
which means that we can directly help them with the core part of the coursework.
Also, since the framework is not optimized for compile-time performance, it is
significantly easier for students to express what they want, especially for
students with less programming experience. Finally, since \toolname{} is written in
Python, which does not need to be compiled, there is no time spent waiting for
recompilation after a single change, and thus students can iterate on their compiler
much faster.

\subsection{Use case 2: Designing a DSL-compiler}
\label{sec:uc2}

\begin{wrapfigure}[9]{r}{13em}
  \vspace{-2em}
  \center
  \includegraphics[width=11em]{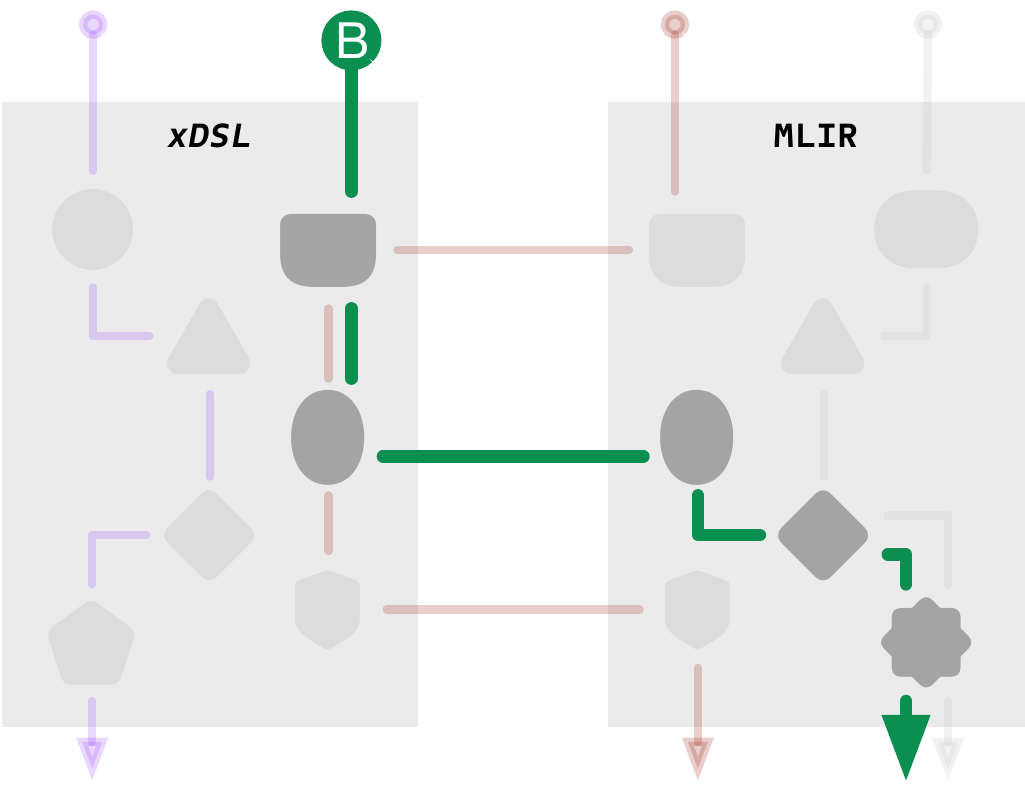}
  \caption{From \toolname{} to MLIR}
\end{wrapfigure}

Many domain-specific compilers reimplement a lot of shared infrastructure for
rewriting, parsing and printing, as well as shared optimizations,
such as loop tiling and vectorization.
Recently, the multi-level rewriting approach has been applied in several contexts showing its potential
(CVM~\cite{CVM} for databases, CIRCT~\cite{CIRCT} for hardware, and
MLIR for general compilation). By modeling every abstraction
level in a distinct way, abstractions, and hence optimizations, can be
shared among multiple compilers across domains. However, getting the
right design for such a compilation stack is challenging, usually requiring several iterations on designs.

We used \toolname{} to develop a compilation stack for the database domain. This
stack has two steps: query compilation and machine-code generation. During query
compilation, queries are optimized on the level of relational algebra and
translated to MLIR IR using \toolname{}. During machine-code generation, MLIR
lowers and optimizes this IR to generate a binary. Our tool can reuse
optimizations across these two steps, which reduces code size compared to traditional systems.

\paragraph{\textbf{User}}

The user is a database engineer, though the concepts apply to many domain
engineers or domain scientists.
Such users can be slowed down by complex build systems and frameworks that require
significant time to get started. Thus, they prefer frameworks that are
easy to familiarize, understand, and allow simple translation of high-level
ideas into code without worrying about low-level implementation details.

For this use case, the most crucial need is a low engineering effort for
defining and modifying dialects, as it allows understanding and exploring
the design space of domain-specific abstractions efficiently. Additionally,
being able to show the performance potential of generated machine code serves
well to convince investors and reviewers of a given design. A minor need is
a fast and easy installation. Even on fast hardware, a slow and complicated
installation process can be a significant hurdle. On the other hand, the need to have
fast compilation speeds is not central for prototyping as, in most domains, the
computation dominates the runtime. 
Furthermore, when working on algorithmic optimizations, users do not
want to be concerned with low-level details like memory allocation, as working on
a high level makes it easier to translate thoughts into code. Python's
simple and high-level code and the ability to work without a complex build system
make it more suitable in this context than low-level languages like C++.

\paragraph{\textbf{Existing Workflows}}

While there is a need to unify domain-specific compilers, they
usually stay in their own domain (e.g., CVM in the database domain) and generate source code (usually C++ or LLVM IR) leveraging general-purpose
compilers to produce binaries. Separate infrastructures make cross-domain
compilation hard and lead to replicated optimizations, e.g. dead code
elimination, or common subexpression elimination.

Using MLIR, one can fuse domain-specific compilers with general-purpose ones
removing the aforementioned code duplication. However, while working with MLIR,
a developer needs to keep low-level details in mind. This complexity incurs a cost
when refactoring code with new design decisions, fixing mistakes, and even
getting started, both on the level of the code (creating a new dialect requires
non-trivial changes in at least two or three files) and on the conceptual
level.

\paragraph{\textbf{The \toolname{} Approach}}

While \toolname{} cannot remove the cost of understanding multi-level rewriting,
its PyPi-based distribution makes it easy to get started.
Additionally, being implemented in native Python makes it more easily usable by
other Python projects. Therefore, users can write in a high-level language without having to think about
low-level implementation details when implementing domain-specific
optimizations. The choice of Python also makes iterating on design
decisions much faster than a C++-based flow.

\begin{figure}
  \centering
  \includegraphics[height=.5\columnwidth]{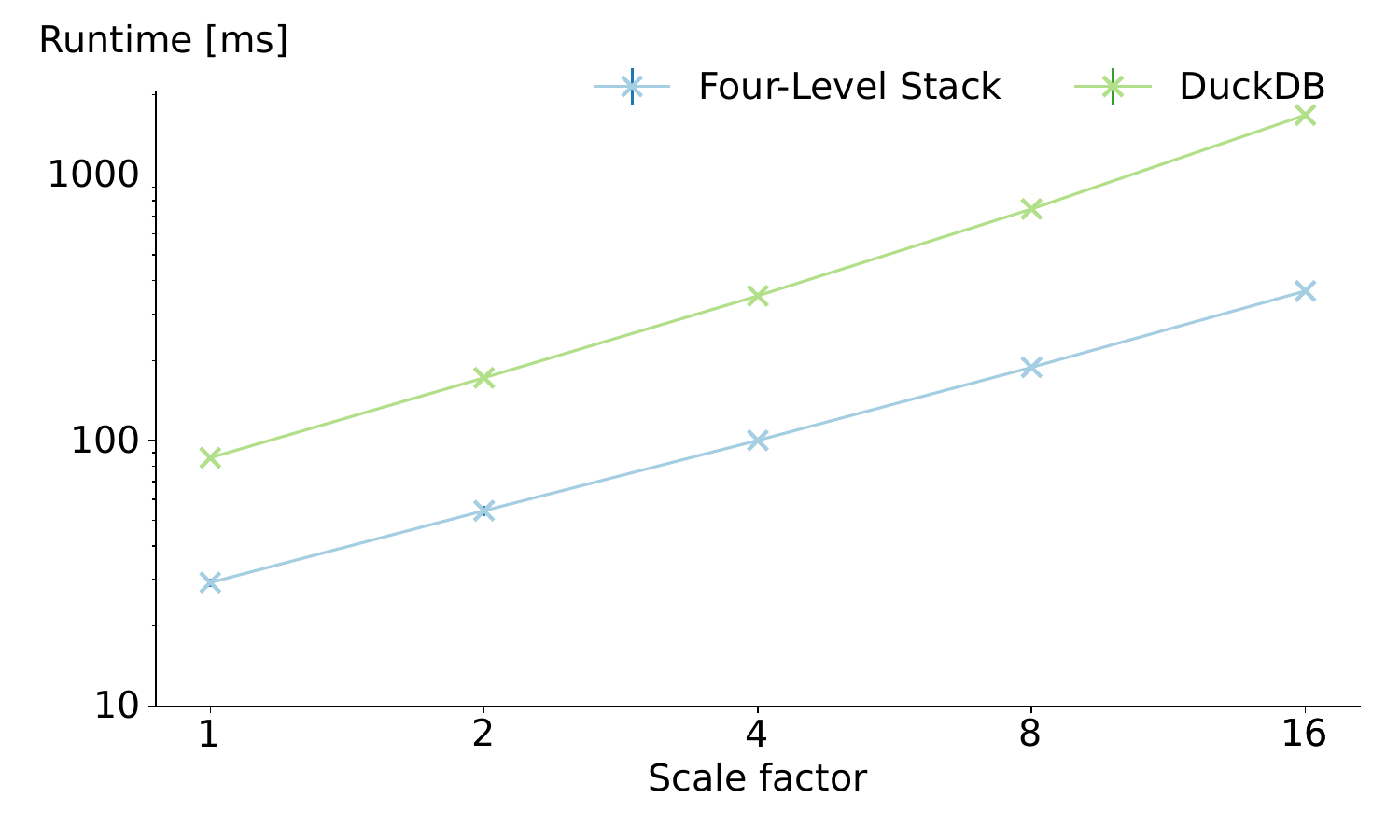}
	\caption{Our Four-Level Stack DSL outperforms the state-of-the-art
	 database DuckDB on the single-threaded Q6 TPC-H benchmark.}
    \label{fig:uc2:runs}
  \vspace{-1em}
\end{figure}

\toolname{}, as a sidekick of MLIR, has access to the 30
industry-quality dialects in MLIR. These are automatically generated through
IRDL, making updates and refactorings on these dialects seamlessly
available after pulling from the MLIR repo. While it is possible to
generate optimized code using \toolname{}, one can also leverage MLIR's, and
hence LLVM's, low-level optimizations by targetting MLIR abstractions. In this
way, our database compiler produce code outperforming
state-of-the-art tools like DuckDB~\cite{DuckDB} by 3x in a single-threaded
setting (\autoref{fig:uc2:runs}). While this workflow needs a compiled version
of MLIR, one does not have to touch the MLIR codebase.

The implemented prototype also applies projection pushdown implemented as four
peephole rewrites to optimize memory accesses for a columnar
database~\cite{databaseBook}. Projection pushdown is a broadly applied
optimization in the database domain that adds as many projections as close to
the base tables as possible. Later on, projections close to the base tables are
fused into the base tables by only loading the strictly necessary columns.

\subsection{Use case 3: Prototyping new MLIR features}
\label{sec:uc3}

\begin{wrapfigure}[11]{r}{13em}
  \vspace{-1em}
  \includegraphics[width=11em]{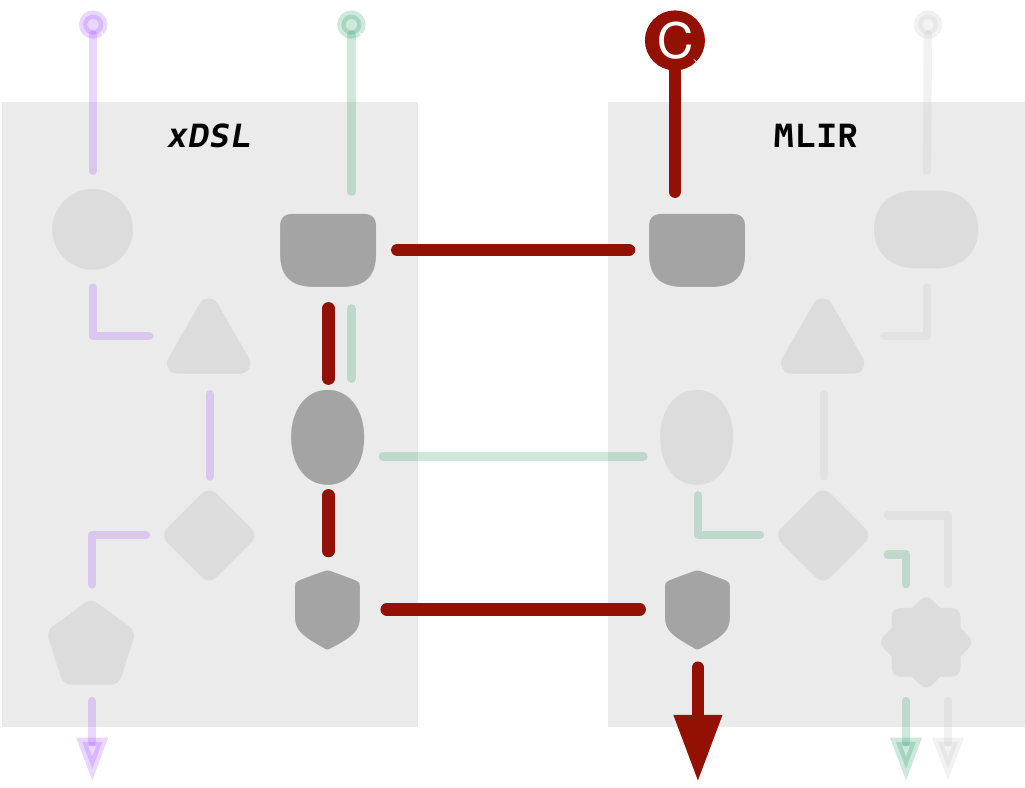}
	\caption{Intercepting the compiler}
\end{wrapfigure}

While most changes to compilers are incremental, such as adding
new transformations, compiler researchers
are also concerned with re-imagining the core design of
compilers. We used \toolname{} to prototype a new approach to a
rewriting system for MLIR, enabling declarative, composable and controllable
rewrites inspired by Elevate~\cite{elevate}. This approach allows for expressing
optimizations using simple abstractions that are composed to form complex rewrites.
The rewriting system leverages an immutable IR, which forbids arbitrary mutations
and enables backtracking with low memory cost.
Making a mutable IR, the de-facto standard, immutable requires invasive modifications in modern compilation
frameworks.

\paragraph{\textbf{User}}
The users are compiler researchers who aim to design the next-generation compilers
by extending or modifying core design aspects. They are familiar with LLVM and
MLIR and understand the codebases well and work individually
or in a small research group using a wide range of hardware, from company-provided
laptops and high-end desktops up to super-computing hardware. They want to prototype
and evaluate new ideas quickly and, if beneficial, publish them and contribute
them back to industry compilers, e.g., via LLVM and MLIR.

\paragraph{\textbf{Needs}}

The number one priority of the users is to prototype ideas quickly and evaluate
their feasibility. Researchers do not want to waste time engineering low-level
details for an idea that is not beneficial. Furthermore, it is essential
for them to iterate quickly and benchmark multiple designs against each
other. Thus, a framework with quick build times is preferred.
While observing the performance and memory trade-offs of different prototypes is
important, it is rarely the aim to achieve production performance
with the prototype. Finally, it is crucial to test a prototype with real-world
programs and observe its results in an end-to-end compilation pipeline. Accordingly,
it is important that a tight integration with MLIR is possible, such as
integrating a prototype into an MLIR pass pipeline.

\paragraph{\textbf{Existing Workflows}}

The existing way to approach a new feature is to experiment with the prototype design directly in MLIR using C++.
This forces the user to split their effort between taking design decisions for a prototype and managing low-level implementation details, such as maintaining existing storage layouts.
These details often have to be revisited later while tweaking the prototype design and arguably put a strain on productivity making it unsuitable for fast iteration.
The burden of long build times on local consumer hardware is amplified when multiple
prototypes have to be designed and compared.
The focus of the MLIR framework leans heavily towards production performance rather than
a clear structure for easy understanding and extensibility of its core concepts. This aggressive optimization for performance
has to be considered constantly and leads to a rigid system with hard-to-understand design
decisions and poor hackability.

We implemented the envisioned rewriting system by investing four months and managed
to design a working prototype of the rewriting system with limited support for
composing rewrites. However, it lacked the foundation of an immutable IR and thus
could not prevent arbitrary IR modifications or support backtracking. For this reason,
it also exhibited several unsolved problems when rewriting complex programs.
While the researcher is very familiar with the C++ programming language and the details of the MLIR
implementation, a large portion of the time was spent on getting the template-based
C++ typing correct with all required features to achieve a composable API interface. Furthermore,
much time was spent implementing features not for the prototype but for tweaking the MLIR framework
and making it usable for this use case. This includes multiple DSLs to ease access to complex
MLIR APIs and combat the verbose nature of C++.

\paragraph{\textbf{The \toolname{} Approach}}

\begin{figure}
  \includegraphics[width=\columnwidth]{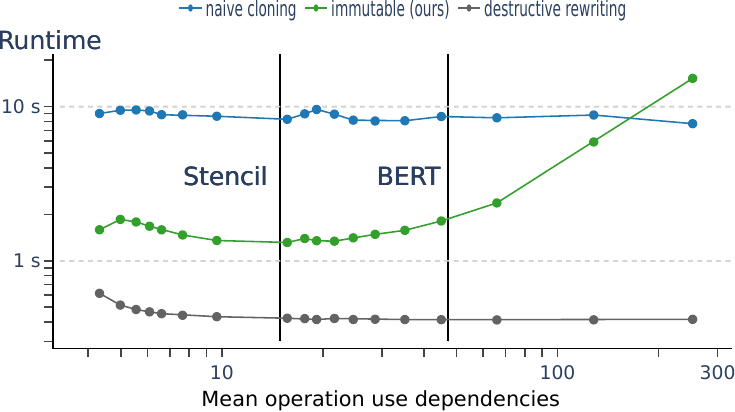}\\
  \caption{Our immutable approach has a runtime between naive cloning and destructive
	rewrites. For code such as climate stencils and BERT, we predict only a 4x
	cost for the benefit of immutability.}
  \label{fig:rewriting_performance}
\end{figure}

\toolname{} exposes the main MLIR concepts and high-level design decisions in a Python interface. Thus, it enables
the implementation and modification of core MLIR constructs while leveraging the productivity of Python.
This makes experimentation with drastically different design approaches in the context of MLIR practical in the first place.
For instance, modifying the core IR infrastructure to support dependent typing can be implemented following a
number of approaches. Instead of modifying numerous C++ files per approach, with little opportunity for
reuse between completely different approaches and handling a complicated build system, in \toolname{}, the core IR can
be flexibly switched by extending or replacing one Python file.
\toolname{} empowers the researcher to switch between different core IR designs using a flag to flexibly benchmark them.

In about 8 weeks, we designed a first working prototype with \toolname{} with equal capabilities to the C++ prototype.
After 4 months, the time required with C++ for a basic prototype, the new prototype had significantly advanced, now backed by an immutable IR infrastructure with full backtracking support.
The similarity in structure to the MLIR framework made the adoption of \toolname{} straightforward without prior experience.
The API interface was easily adjustable to the current needs.
With fewer requirements for managing low-level implementation details,
it became possible to focus solely on the design of the prototype and iterate quickly.
The close interaction of \toolname{} and MLIR made benchmarking the rewriting of real-world
machine learning models practical.
Our prototype shows the advantages of leveraging an immutable IR for a backtracking
rewriting system and allows us to evaluate the trade-off in memory and processing speed
when using a representation of IR that can efficiently recover a previous state in the
compilation process.
\autoref{fig:rewriting_performance} evaluates the approaches' rewriting time and memory
consumption by varying the IR structure to be optimized, i.e., the number of uses of the mean operation.
A mutable rewrite system that performs a naive cloning of the IR to support the backtracking pays a high overhead independent of the IR structure.
In contrast, the performance of our immutable rewriting system heavily depends on the structure of the IR.
The more uses the mean operation has, the less reused; hence, rewriting time and memory consumption increase.
However, for the structure of real-world use cases, such as weather modeling stencil computations from the Open Earth Compiler~\cite{gysi2021domain} (left vertical line)
or the BERT-small~\cite{devlin2018bert} (right vertical line) transformer model, our approach requires less rewriting time and consumes much less memory.

\section{Instantiating Core Compilation Concepts}
\label{sec:pykick}
Identifying the compiler framework to connect to and the core
compilation concepts to expose is a core step in implementing a sidekick
compiler. MLIR, the framework \toolname{} connects to, demonstrated that pairing
SSA with block arguments, nested regions, and attributes yields an effective set of core abstractions.
\toolname{} tailors the implementation of SSA and multi-level rewriting to Python, in a
manner differing from MLIR's C++ and performance-focused implementation.
We now describe the data structures and concepts
that are shared by both \toolname{} and MLIR.

\subsection{Operations}

Operations are the core structure of the IR and represent both computations and
control flow structures (e.g., loops). An operation consists of a name, a list
of operands, a list of results, a dictionary of attributes
(\autoref{sec:ir-attributes}), a list of regions, and a list of successors
(\autoref{sec:ir-regions}). When defining an operation, we also provide a set of
invariants over these structures, called the verifier, which checks that the operation has
the correct input and results. For instance, the \texttt{arith.addi} operation
expects two operands and a single result, all with matching integer types. Each
operation also defines informally its own semantics (e.g., in its written
documentation). Here, \texttt{arith.addi} returns the addition of the runtime
value of its inputs. Operations have a generic textual format (upper line in the
following example) but can also be extended with a custom format (lower line)
for conciseness and readability:

\begin{mlir}
            : (i32, i32) -> i32
\end{mlir}

\subsection{SSA Values}

In SSA-based IRs, operations are connected through SSA
values, which represent the runtime values of objects.
Each SSA value has a single, statically known, definition, either as a result of
an operation or as a block argument.
Each value is then used zero or more times, as an operand to operations.
The graph linking definitions and uses is what is traditionally called the def-use
chain. The SSA property makes this graph simple, making analysis less expensive.
Having SSA as a core property of the IR gives a unified structure that allows
representing and manipulating IRs in a generic way.

\subsection{Attributes and Types}
\label{sec:ir-attributes}

Attributes encode compile-time information (e.g., constants or types) and are
used to parametrize both SSA values and operations. Types are special
attributes that provide static constraints over SSA values. For instance, the \texttt{i32} type
on a value encodes a signless 32-bit integer. Attributes attached to
operations through their attribute dictionary encode operation parameters. For
instance, the operation \texttt{arith.constant} with the named attribute
\texttt{value = 42\,:\,i32} encodes that the resulting SSA value is the integer
42 encoded in 32 bits:

\begin{mlir}
           : () -> i32
\end{mlir}

In the textual format, attributes start with \texttt{\#}, and types start with
\texttt{!}. Builtin MLIR attributes that are frequently used have a shorthand notation, such as \texttt{i32}.
Both attributes and types can be parametrized with arbitrary data. For
instance, \texttt{!complex<i32>} is the \texttt{complex} type parametrized with
the \texttt{i32} type, and \texttt{[2\,:\,i32, 5\,:\,i32]} is the builtin
\texttt{ArrayAttr} attribute parametrized with the \texttt{2\,:\,i32} and
\texttt{5\,:\,i32} attributes.

This definition of attributes and types differs slightly from their
original MLIR meaning. In MLIR, attributes and types are entirely disjoint,
though there exists a \texttt{TypeAttr} attribute that encodes a type as an
attribute.
We made types a special case of attributes to simplify the design of the core language.
Without the constraint of wanting to support MLIR syntax, we would not have defined types
at all and allowed the use of any attribute as a type annotation.

\subsection{Blocks and regions}
\label{sec:ir-regions}

While operations represent computation and structures, they are insufficient to
represent control flow graphs. Instead, structures that reason about control
flow in a larger sense are needed. In \toolname{}, the corresponding structures
are blocks and regions.

A block is a sequence of operations that execute in order. Arranging blocks into
a (potentially cyclic) directed graph yields a Control Flow Graph (CFG). Each
CFG has a single input block that is executed first, but no specific termination
block, and the control flow may leave the CFG at any block's final operation. To
that end, these final operations are constrained to be terminator operations, which are a
subset of operations that must specify which successor blocks can be jumped to.
This restriction can however be lifted for single-block CFGs. Furthermore,
blocks may have block arguments that introduce SSA values, giving the IR a
functional structure that has been shown to be equivalent to
phi-nodes~\cite{GlobalValueNumbering}.
\vspace{-.5em}
\begin{mlir}
  ^block0(
    scf.cond_br 
  ^block1: ...
  ^block2: ...
\end{mlir}
\vspace{-.5em}

A CFG is placed in a region, which is nested in an operation. In
contrast to IRs like LLVM's, which require analysis passes~\cite{llvmLoops},
regions model nested control flow as first-class
constructs to represent structures such as loops or conditionals. For instance,
a conditional can be represented with a \texttt{scf.if}, which uses regions to
encode both execution branches. The operation that contains each region defines when that
region is executed, and control flow is returned to this operation
when the region terminates. For instance, the regions in a \texttt{scf.if} are
conditionally executed on the operation operands, whereas the ones in a
\texttt{scf.for} can be executed multiple times.

\vspace{-.5em}
\begin{mlir}
  scf.if 
    // True region
  } else {
    // False region
  }
  \end{mlir}
  \vspace{-1em}

\subsection{Dialects}

Operations and attributes that represent similar concepts are grouped in
dialects, allowing separation of concerns. For instance, the \texttt{arith}
dialect contains operations for simple arithmetic, and the \texttt{scf} dialect
for operations with structured control flow, such as loops and conditionals.
This separation of concerns allows multi-level compilation pipelines, which
interleave domain-specific optimizations of dialects with progressive
lowerings to lower-level dialects. Hence, multiple dialects can be used in
the same program, allowing lowerings to only target parts of a program.

\subsection{MLIR Compatibility}

We designed \toolname{} IR to be compatible with MLIR IRs to simplify interactions
between the two frameworks. Notably, it allows \toolname{} and MLIR to share the
same textual representation, which allows transferring programs directly through
parsing and printing. This interoperability is essential for a sidekick framework,
allowing to use one framework as a drop-in replacement of the other, letting the
user decide which tool is best suited for their current needs. Additionally,
it simplifies the transition of code from one framework to the other, since they
both use the same IR structure and concepts.

Since operations have the same generic format in both frameworks, they can be
automatically translated from one to the other. One can even allow unregistered
operations, meaning operations without a definition in the framework, to be parsed and
printed. However, while unregistered attributes can be parsed and printed,
they cannot be freely manipulated, as they are only represented with their string representation
since attributes lack a generic format.

\section{Defining and Sharing Dialect Definitions}
\label{sec:dialect-definition}

While both \toolname{} and MLIR define dialects using their respective
programming languages, efforts have been made in the MLIR project to make these
abstractions more declarative through the IRDL language \cite{irdl}.
We leverage this work with the IRDL dialect, which uses an SSA
representation to define MLIR abstractions as programs, allowing us to share one
abstraction definition between \toolname{} and MLIR. We also provide a frontend for IRDL
in Python to provide a better interface for defining abstractions in
\toolname{}.

\subsection{The IRDL dialect: An IR for IR definitions}
\label{sec:irdl-ssa}

The IRDL dialect is defined in MLIR and \toolname{}, allowing both
frameworks to share dialect definitions the same way they share programs.
Programs using the IRDL dialect define dialects, types, attributes, and operations using a
small but expressive constraint engine inspired from IRDL. Making IRDL a dialect
allows one to easily embed it in any compiler framework offering SSA dialect
infrastructure, such as MLIR or \toolname{}. As dialects are now input data
for compiler infrastructures, IRDL-dialect-defined dialects inherit the
introspectable, portable and transformable nature of any other IR program.

\begin{figure}[b]
    \begin{mlir}
Dialect cmath {
  Type complex {
    Parameters (elem: !AnyOf<f32, f64>) } }
    \end{mlir}
    \vspace{-.5em}
    \centering $\downarrow$
    \begin{mlir}
irdl.dialect @cmath {
  irdl.type @complex {
    irdl.parameters(
} }
    \end{mlir}
  \caption{Our new IRDL dialect exposes the IRDL language definitions (top) as
	an IR program (bottom), which can be easily shared across compilers.}
  \label{lst:irdl-ssa-example}
\end{figure}

\begin{figure*}
  \includegraphics[width=\textwidth]{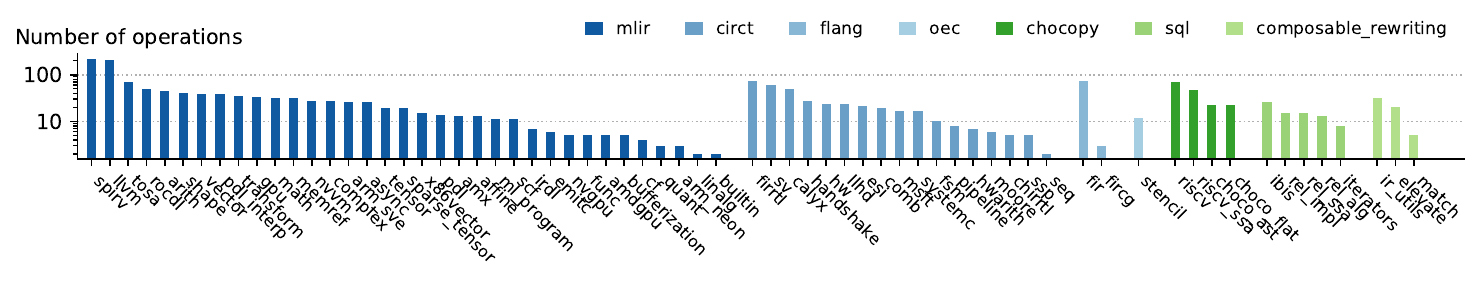}
    \caption{\toolname{} combines its 12 SSA dialects (\nTotalToolNameOps{}~ops) (green) with the
    54 SSA dialects (\nTotalMlirOps ops) defined in the Flang, CIRCT, and MLIR
    Core community repositories (blue).}
  \label{fig:dialects}
  \vspace{-.5em}
\end{figure*}
Dialects, operations, types, and attributes are defined using the \texttt{irdl.dialect},
\texttt{irdl.operation}, \texttt{irdl.type}, and \texttt{irdl.attribute} operations (\autoref{lst:irdl-ssa-example}).
Attributes are defined by constraining their parameters, and operations are
defined with constraints on their operands, results, attributes, successors,
and regions. In the IRDL dialect, each SSA value represents an attribute (or a type),
and constraints over these attributes are expressed through IRDL dialect operations.
For instance, \texttt{irdl.is}, constrains an attribute to be equal to the given
attribute, and \texttt{irdl.any\_of}, constrains an attribute to be one of
the given attributes.

We are working with the MLIR maintainers to bring the IRDL dialect to the main MLIR repository.
The current implementation can register new IRDL dialects at
MLIR runtime, but does not yet support generating C++ definitions. Also, while the IRDL dialect
already provides a way to register operations and attributes that can be defined
purely declaratively, it does not yet provide escape hatches to the framework
language for more complex constraints.

\paragraph{An implementation-agnostic concept}

The sidekick compiler approach is challenging to put into practice when dialect
definitions are deeply embedded within one compiler. Instead, the IRDL dialect
represents dialect definitions in a compiler-agnostic manner. All sidekick
compilers implementing an IRDL dialect-like registration endpoint can easily
share dialect definitions with each other, as long as they can translate
their dialect definitions to the IRDL dialect. This approach allows sharing
the core concepts of the modeled dialects without having to reimplement them
in each compiler.

\begin{figure}[b]
  \begin{pythoncode}
    T = TypeVar("T", bound=Union[f32, f64])
    @irdl_attr_definition
    class ComplexType(Generic[T], ParametrizedAttribute):
      name = "cmath.complex"
      elem: ParameterDef[T]
  \end{pythoncode}
\caption{Defining dialects using a Python EDSL allows us to translate them back and forth to IRDL.}
\label{lst:pyrdl-attr}
\end{figure}

\subsection{PyRDL: Connecting IRDL to \toolname{}}
\label{sec:pyrdl}

\toolname{} dialects are registered using PyRDL, a Python EDSL (Embedded DSL)
inspired by IRDL. An example is shown in \autoref{lst:pyrdl-attr}. PyRDL defines accessors, verifiers, and parser / printer
functions for types, attributes, and operations. The EDSL is also type-safe, so
Python type-checking tools will correctly understand the types of operand
or attribute definitions. For instance, the Python type \texttt{ComplexType[f32]} is the attribute
instantiated by \texttt{!cmath.complex<f32>}.

Operation and parametrized attribute definitions can be automatically translated
back and forth to the IRDL dialect. The translation to the IRDL dialect is done
using Python introspection, while the translation from the IRDL dialect to PyRDL
is implemented as a Python script. In particular, Data attributes cannot be translated
to the IRDL dialect, as they rely on user-defined Python data structures that cannot
be understood by the declarative nature of the IRDL dialect. Similarly, C++ constraints
and attributes with C++ parameters can be translated to PyRDL, but with a generic
of any value.

\begin{figure}[h]
  \hspace{-1em}
  \includegraphics[width=\columnwidth]{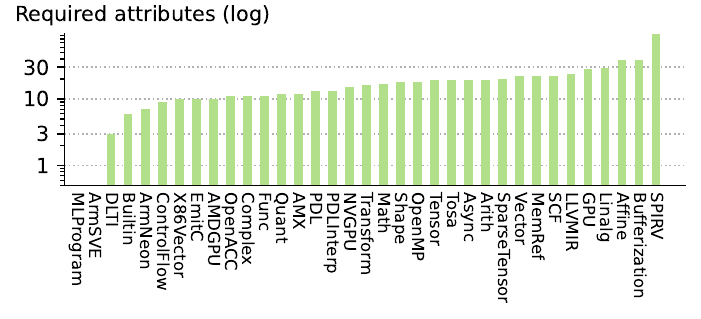}
  \caption{The full use of a dialect in \toolname{} requires few attributes.}
    \label{fig:required-attrs-per-dialect}
  \vspace{-1em}
\end{figure}

\subsection{A shared dialect ecosystem}

Using the translation from the IRDL dialect to PyRDL, we can use the entire MLIR
ecosystem in \toolname{}. Combining the projects using MLIR, such as MLIR,
Flang, or CIRCT, with several \toolname{}-based projects, leads to an ecosystem
with a plethora of dialects and operations (\autoref{fig:dialects}). While the
number and name of operands, results, attributes, successors, and regions are
always translated, the constraints defined in C++ rather than in IRDL are not,
and instead are replaced by a generic constraint that accepts any value.

While operations can be translated from one framework to the other using
the operation generic format, attributes (and thus also types) do not
have generic format. MLIR does provide a declarative specification for
attribute custom format, but it is not used by every type and attribute,
and may embed arbitrary C++. Thus, to convert attributes used in programs,
we need to manually implement a printer and a parser for each attribute
definition. To quantify the number of attributes that need to be ported to
use a dialect fully, we count the number of distinct attributes used in the
dialect test folder in the MLIR test suite (\autoref{fig:required-attrs-per-dialect}).
We find that most dialects only use a few distinct attributes and most dialects
require fewer than ten attributes to be fully usable in \toolname{}.
Additionally, all dialects but the SPIRV dialect require fewer than 30 attribute
definitions.

\section{Characterization of the compiler design space}
\label{sec:design_space}

To better understand the design-space trade-offs between \toolname{} and more
traditional production frameworks, we compare against MLIR on multiple metrics
relevant during compiler development. To that end, we not only compare
both compilers on their runtime but also on the time taken to compile and
install the compilers themselves. This comparison shows that while \toolname{}
has a slower runtime than MLIR for larger files, it runs in the same order
of magnitude as MLIR for smaller files, such as those used for testing.
Furthermore, we show that \toolname{} uses significantly fewer resources
to install and run after a change in the compiler. This comparison demonstrates
that \toolname{} offers users a point in the compiler design space that
prioritizes developer productivity over the performance of the compiler
executable.

\subsection{Startup time and build time}

Two important metrics that are often overlooked in a compiler framework are
the costs of installing and running it for the first time, and
running it after an incremental change. While these features rarely matter for
end users, they do for compiler developers. We
compare these metrics between \toolname{} and MLIR, both compiled in release
mode with assertions enabled, which enables compiler optimizations, and in debug mode, which does not enable
optimizations and allows the use of debuggers. Both versions are compiled with
Clang version 14.0. We compare these metrics on two devices
that correspond to two potential users: A desktop using an AMD Ryzen 9
5950X 16-Core CPU, which is a relatively high-end CPU that could typically
be used by a compiler engineer, and a middle-end laptop with an Intel i5
10210U 4-Core CPU.

\begin{figure}[h]
  \hspace{-.8em}
  \begin{subfigure}[b]{0.23\textwidth}
    \includegraphics[width=\textwidth]{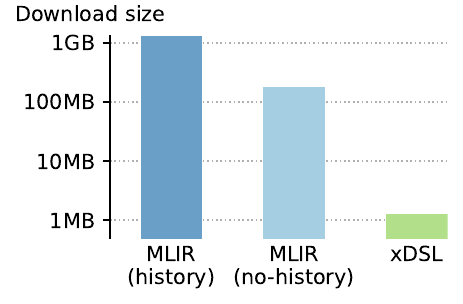}
  \end{subfigure}
  \hspace{-.2em}
  \begin{subfigure}[b]{0.23\textwidth}
    \includegraphics[width=\textwidth]{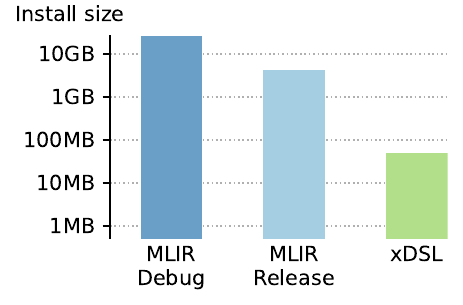}
  \end{subfigure}\\
  \begin{subfigure}[b]{0.23\textwidth}
    \includegraphics[width=\textwidth]{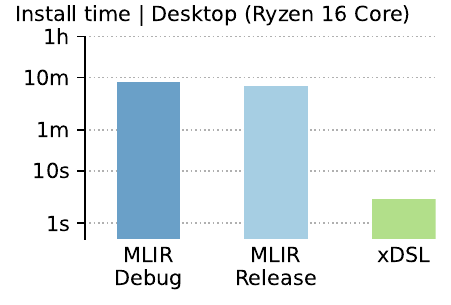}
  \end{subfigure}
  \begin{subfigure}[b]{0.23\textwidth}
    \includegraphics[width=\textwidth]{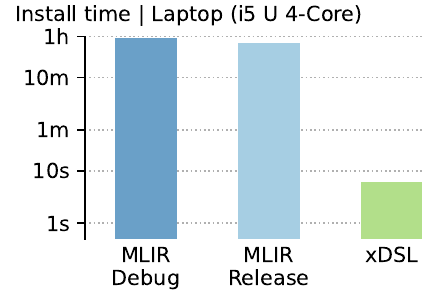}
  \end{subfigure}\\
	\caption{\toolname{} installation is lightweight, making it usable on low-end machines.}
  \label{fig:startup-time}
  \label{fig:install-time}
\end{figure}

\paragraph{Startup time}
First, we compare the startup time, i.e. the time necessary to run the test
suites for the first time after downloading the repository. This
is representative of the use of MLIR and \toolname{} since it is the only
way to modify the abstractions and passes included in the projects.
While installing \toolname{} is done by a local install with
\texttt{pip}, installing MLIR is done by compiling it with \texttt{CMake}.
On both machines, compiling MLIR requires two orders of magnitude more time
than \toolname{} (\autoref{fig:startup-time}), taking almost 1 hour on the
laptop and 10 minutes on the desktop, compared to the few seconds the \toolname{}
setup needs on both machines. We observe that the compilation on the laptop is
significantly slower than on the desktop as the laptop only has a few cores, and
the compilation of MLIR can be heavily parallelized.

\paragraph{Download and installation size}
Most users will install the projects manually on their machines, requiring
to install around 2MB for \toolname{}, and 1GB for MLIR (which can be lowered
to 100MB by doing a shallow git clone). However, some users will
instead opt for a pre-compiled version of MLIR, for instance compiled in a
docker container. While this reduces the cost of compiling MLIR, it still
requires significant disk space. While the entire folder of \toolname{},
including a virtual environment for Python, requires less than 50 MB, the
install folder of MLIR is more than 4GB in release mode and 26GB in debug mode
(\autoref{fig:install-time}), resulting in long download times for MLIR
binaries. Note that the installation size of MLIR is lower for users
that only need a subset of the dialects or executables. \toolname{}' smaller
installation size simplifies its distribution for many users and enables its
use in cloud or web environments.

\begin{figure}
  \includegraphics[width=\columnwidth]{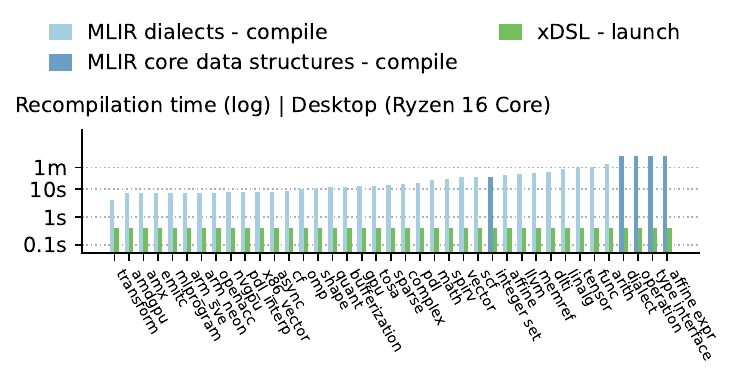}

  \caption{Recompiling MLIR after a single change is up to two orders of magnitude slower
           than launching \toolname{}, impacting prototyping and development efficiency.}
  \label{fig:recompilation-mlir}
\end{figure}

\paragraph{Incremental builds}
Another interesting metric is the time taken to launch the compiler after a single change
in its definition. This metric represents the daily use of the frameworks,
as having a lower recompilation time is essential for rapid iteration on changes in the
compiler. To measure this, we add a single space character at the end of a
file containing a dialect or core data structure implementation and measure the time to
launch (\autoref{fig:recompilation-mlir}).
While MLIR requires a costly partial recompilation after a change, \toolname{}
startup time doesn't change, being the cost of loading the Python source files.
On the desktop, the majority of dialects require over than 14 seconds of
recompilation, up to a minute, and some core data structures require more
than 2 minutes of recompilation after a change.
This is worse on the laptop, where more than 40\% of the dialects require a
recompilation of more than a minute, up to 10 minutes. When iterating on
dialects on low-to-middle-end hardware, MLIR recompilation times are
significant, slowing down developers substantially. On the other hand,
\toolname{} is removing that friction on both high and low-end hardware.

\subsection{Runtime performance of \toolname{}}

We measure the compile-time performance of \toolname{} and MLIR by comparing
the parsing and printing time of the MLIR test suite, as well as the runtime
of a simple constant folding pass on large synthetic files. We compare the
runtime performance of MLIR compiled in release and debug mode against the
runtime of \toolname{}, both from a new Python subprocess, and from a
tool that already preloaded all available dialects. The reason for this is
that the Python decorators used in PyRDL (\autoref{sec:pyrdl}) have a fixed
cost when running multiple files from the same Python script, and this
cost is currently the bottleneck when running small tests, compared to MLIR,
which is optimized to have a fast loading time.

\begin{figure}[h]
  \hspace{-1em}
  \includegraphics[width=\columnwidth]{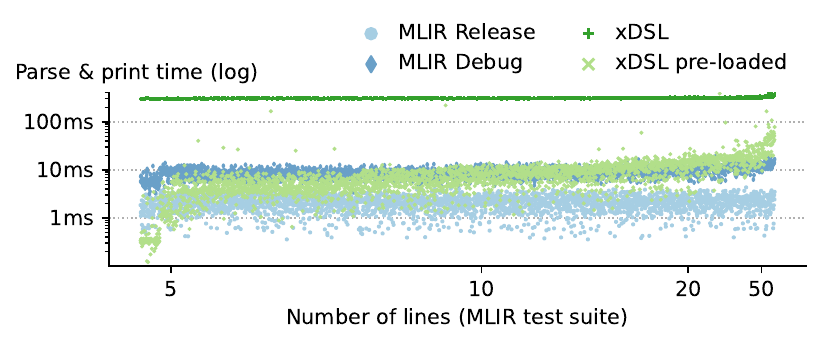}

  \caption{A Pre-loaded \toolname{} parses and prints production
           test cases at a similar speed as MLIR compiled in Debug mode, making it
           a suitable alternative for testing a prototype.}
  \label{fig:parsing-time}
\end{figure}

\begin{figure}[h]
  \hspace{-1em}
  \includegraphics[width=.95\columnwidth]{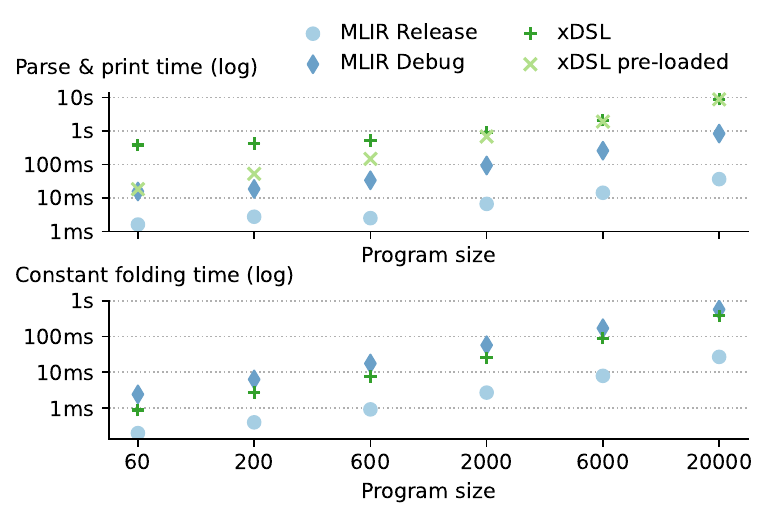}
  \caption{While parsing and printing large files in \toolname{}
  is an order of magnitude slower, the introspection and modification
  of IR is comparable to MLIR compiled in Debug mode.}
  \label{fig:parse-print-constant-folding}
\end{figure}

\paragraph{Parsing and printing}
We compare the parsing and printing time between \toolname{} and MLIR by parsing and
printing all files in the MLIR test suite, as well as large synthetic files.
We parse and print \percentageTestsInFoldersPassed{} of the \nTestsInFolders{}
programs in the MLIR test suite (\autoref{fig:parsing-time}), where failing tests are due to our partial
implementation of the \texttt{builtin} attribute parsers and printers.
Overall, we observe that MLIR compiled in Debug mode, which is
the standard way of using MLIR during prototyping, takes
\suiteParsingTimeMLIRDebug{}s to parse and print all tests.
When compiled in Release mode, the same benchmark takes \suiteParsingTimeMLIRRelease{}s.
\toolname{} is more than an order of magnitude slower to parse the same files, and takes
\suiteParsingTimexDSL{}s.
This is due to the cost of launching the interpreter and importing \toolname{} for every test.
If all the files are processed in the same interpreter context, \toolname{} takes
\suiteParsingTimexDSLNative{}s, which is roughly similar to MLIR compiled in Debug mode.
When parsing and printing large files
(\autoref{fig:parse-print-constant-folding}), we observe that the same trend holds,
besides that Python overhead is not significant anymore, and thus removing it
does not provide any benefits.
Note that \toolname{} does not define all dialects used in the MLIR test suite, and that
some attributes and types are parsed with a default parser that only checks that the attribute has a balanced set of brackets.

\paragraph{Runtime}
We evaluate the runtime performance of both compilers by defining a constant
folding pass, which rewrites additions of constants to a single constant. As
this is a simple pass, it requires almost no logic, such that most of the
runtime is spent in the compiler framework for introspecting the IR or
modifying it. We run this pass on large test cases, such that the rewrite is
applied on each possible operation, to show how the performance of
\toolname{} scales compared to MLIR (\autoref{fig:parse-print-constant-folding}). Overall, the runtime of the pass is
around twice as fast as the same pass in MLIR compiled in debug mode
and is around an order of magnitude slower than MLIR compiled in release mode.
While MLIR compiled in release mode is clearly preferable when runtime speed is a concern,
\toolname{} has comparable speed to MLIR in a debugging environment.

\section{Related Work}

With sidekick frameworks, we explore new ways of making compilers and compiler
frameworks from different communities interact. This section describes other
approaches compilers have used to connect with compilers from other communities.
We also look at other compilers implemented in Python, which share common design
goals with \toolname{}.

\paragraph{Textual interoperation with LLVM}
As LLVM defines a stable textual IR, multiple tools have been developed using
it. For instance, Alive~\cite{10.1145/2737924.2737965} is a popular tool that
uses LLVM IR textual format to do translation validation for LLVM. Similarly,
Vellvm~\cite{10.1145/2103621.2103709} uses the LLVM IR textual format to provide
mechanized formal semantics for LLVM IR. In particular, Vellvm implements
a verified version of the LLVM mem2reg~\cite{zhao2013formal} pass that can be
used at any point during a compilation pipeline through the use of LLVM IR
textual format. While Vellvm connects loosely with LLVM using the textual format,
this connection is not extensible, and not as automatic as in the sidekick
approach we propose.

\paragraph{Language bindings}
Popular compiler frameworks provide APIs for other languages,
such as Python, showing the need to interact with compilers through a high-level
scripting language. For instance, llvmlite provides Python bindings
for LLVM~\cite{LLVMlite}, and MLIR has Python bindings~\cite{MLIRPythonBindings}
in its main repository which is notably used by Nelli~\cite{levental2023nelli}
to define an eDSL for writing MLIR programs. MLIR's Python bindings expose IR constructs like
dialects, operations, regions, and attributes, but also a pass manager,
enabling an entire compilation flow through Python.
However, the Python bindings are still bindings to a C++ framework. So, if one
intends to modify or create a new dialect, or to change the behavior of MLIR,
say by adding a new rewriting infrastructure, one still has to work on the
C++ code. Furthermore, to interact with MLIR dialects that are not exposed
yet, such as the LLVM IR dialect, the corresponding Python bindings must
be added manually. In contrast, \toolname{} allows extending its dialects
directly in Python while also allowing to export MLIR dialects. Also,
\toolname{} is completely written in Python, so changing the framework
does not require any other language.

\paragraph{Lowerings to other compilers}
Similar to how \toolname{} can be used to interact and generate machine code
through MLIR and LLVM, many compilers generate an intermediate representation
and hand it to a middle-end compiler to generate machine
code. For instance, Clang~\cite{lattner2008llvm},
the Rust~\cite{klabnik2019rust} compiler, and the
Julia~\cite{bezanson2017julia} compiler generate LLVM IR and uses LLVM
to generate machine code, allowing them to target a wide range of architectures.
CVM~\cite{CVM}, a framework for multi-level rewriting in the
database domain uses an infrastructure written in Python to define its high-level IRs,
and then generates data structures of state-of-the-art execution
layers like MonetDB~\cite{MonetDB} to execute the code.
Though these pipelines
have interactions between compilers and sometimes from different languages,
they do not have the bi-directionality of \toolname{} and MLIR, which allows to
go back and forth during the different stages of the compilation pipeline.

\paragraph{Compilers in Python}
Other compilers have been implemented in Python, though
these compilers are often only targetting Python itself. For instance,
Nuitka~\cite{nuitka} an ahead-of-time Python compiler that targets C
code, Numba~\cite{lam2015numba}, a just-in-time compiler for Python that
targets LLVM IR, and Scalpel~\cite{li2022scalpel}, a static analysis
framework for Python. More recently, the PyTorch~\cite{NEURIPS2019_9015}
community introduced \texttt{torch.fx}~\cite{reed2022torch}, a framework
to define transformations for PyTorch kernel directly in Python. While
these compilers are written in Python to make them accessible to their
community of users, they are restricted to their domain and are
not extensible as \toolname{} and MLIR are.

\paragraph{Extensible compilers}
Other compiler frameworks have similar concepts of extensibility as the heavy
linking of \toolname{} with MLIR. For instance JastAdd~\cite{JastAdd}, Graal
IR~\cite{duboscq2013graal}, both in Java, and Delite~\cite{sujeeth2014delite},
built on top of Lightweight modular staging~\cite{rompf2010lightweight} in
Scala, are compilers that allow to extend the IR with new abstractions. While we
could have built \toolname{} as a sidekick of these compilers, we decided on
MLIR as the base framework due to the range of abstractions it supports,
compared to the IRs of these frameworks that often lack concepts such as
regions, block arguments, or attributes. On the other hand, the nanopass
framework \cite{sarkar2005educational}, which is an extensible compiler
framework written in Scheme, is a good example of a framework exploring a design
space of quick prototyping and ease of use. Compared to our work, it
does not connect to an industry-ready compiler, and cannot easily be used to
prototype for an existing compiler.

\section{Conclusion}

We introduced the concept of a sidekick compiler framework, a compiler framework
that is loosely coupled to a base framework through shared core concepts while
offering a community-tailored implementation that instantiates these concepts.
We implemented our idea by developing \toolname{}, a Python-native sidekick to
MLIR and demonstrated how \toolname{} instantiates the concepts of SSA, regions,
and multi-level rewriting. Subsequently, we showed how the IRDL dialect enables
us to represent IR definitions as compiler IRs and how this enables the exchange
of both IR definitions and programs across compiler frameworks. Our evaluation
demonstrated that an order of magnitude speedup for incremental builds and a
significantly smaller installation time, making \toolname{} a great tool for
interactive compiler development. We leverage this advantage in three case
studies and showed that sidekick compilers can benefit various developer
communities outside of the classical userbase of the base compiler. We also
demonstrated that by expanding the MLIR ecosystem to a different design space,
we can interconnect communities such that the base and the sidekick framework
can form a single ecosystem of compiler abstractions and transformations.

While the \toolname{} is tailored to the Python ecosystem where interactive
development is a focus, we believe that sidekicks can be built for other
communities or needs.  For instance, one could imagine a sidekick tailored for
portability in Java or Scala, a sidekick focused on parallelism or distributed
computing in Rust, or a sidekick focused on providing correctness guarantees in
Lean or Coq. While a change in programming language can guide the choice of a
new focus, this change is not fundamentally required to explore a new design
space. Similarly, sidekicks implemented in different languages but with similar
technical decisions can be used to connect different programming language
communities. We envision that the translation approach of \toolname{}, i.e.
exposing data structure definitions as programs that can be ported across
frameworks, will be a cornerstone of a new generation of compiler engineering
that heavily leverages sidekick frameworks.

\section{Acknowledgments}

This work was supported by the Engineering and Physical Sciences Research Council (EPSRC) grants EP/W007789/1 and EP/W007940/1.
This project has also received funding from the European Union’s Horizon EUROPE research and innovation program under grant agreement no. 101070375 (CONVOLVE).
Authors would like to thank all of the xDSL community for their useful comments and discussions along the way.

\newpage

\bibliography{references}

\end{document}